\documentclass[AMA,STIX1COL]{WileyNJD-v2}
\usepackage{algorithm, algpseudocode}
\usepackage[nolist]{acronym}
\newcommand{\nfiles}{n}
\newcommand{\npcached}[1]{w_{#1}}

\newcommand{\npacket}{k}
\newcommand{\codedpacket}{c}

\newcommand{\indexi}{i}
\newcommand{\coeff}{g}

\newcommand{\radius}[1]{r_{#1}}

\newcommand{\dcenter}{d}
\newcommand{\memory}{M}

\newcommand{\Pfail}{P_F}

\newcommand{\lowerb}[1]{l{#1}}
\newcommand{\upperb}[1]{u{#1}}

\newcommand{\Pro}{P}

\newcommand{\Psuc}[1]{P_{#1}} 
\newcommand{\library}{\mathcal{F}}
\newcommand{\libraryf}[1]{\mathsf{f}_{#1}}

\newcommand{\hubs}{h}

\newcommand{\probHub}[1]{\gamma_{#1}}
\newcommand{\probFile}[1]{\theta_{#1}}

\newcommand{\probFail}[1]{P_f{#1}}

\newcommand{\isymb}{u}
\newcommand{\isymbv}{\mathbf{\isymb}}

\newcommand{\rosymb}{y}
\newcommand{\rosymbv}{\mathbf{\rosymb}}
\newcommand{\G}{\mathbf{G}}
\newcommand{\Grx}{\tilde{ \mathbf{G}}}
\newcommand{\Expd} [1]{\mathbb{E}[\Delta_{#1}] }

\newcommand{\er}{\textcolor{black}}

\newcommand{\reviewerone}{\textcolor{black}}
\newcommand{\reviewertwo}{\textcolor{black}}

\newcommand{\fran}{\textcolor{black}}
\begin{acronym}

\acro{LRFC}{linear random fountain code} 
\acro{GEO}{geostationary Earth orbit}
\acro{ILP}{Integer Linear Programming}
\acro{LP}{Linear Programming}
\acro{MDS}{Maximum distance separable}
\acro{HAP}{High Altitude Platform}

\end{acronym}

\articletype{Article Type}%

\received{}
\revised{}
\accepted{}
        
\begin{document}

\title{Caching in  Heterogeneous Satellite Networks with Fountain Codes}

\author{Estefan\'ia Recayte}
\author{Francisco L\'azaro}
\author{Gianluigi Liva}

\address{Institute of Communication and Navigation, DLR - German Aerospace Center, Wessling, 82234 Germany \thanks{"Caching in Heterogeneous Satellite Networks with Fountain Codes", E. Recayte, F. L\'azaro, G. Liva, International Journal of Satellite Communications and Networking, Special Issue, John Wiley Sons, Ltd   Copyright  \copyright  2019} 
\authormark{Recayte \textsc{et al} }}

\corres{*Estefan\'ia Recayte \email{estefania.recayte@dlr.de}}

\abstract[Summary]{In this paper we investigate the performance of caching schemes based on fountain codes in a  heterogeneous satellite network. We consider  multiple cache-aided hubs which are connected to a geostationary satellite through backhaul links. With the aim of reducing the average number of transmissions over the satellite backhaul link, we propose the use of a caching scheme based on fountain codes. 
We derive a simple analytical expression of the average backhaul transmission rate and provide a tight upper bound on it. \fran{Furthermore, we show how the performance of the fountain code based caching scheme is similar to that of a caching scheme based on maximum distance separable codes.}
}


\maketitle

\section{Introduction}

Cache-aided delivery protocols represent a  promising solution to counteract the dramatic increase in demand for multimedia content in wireless networks.  Caching techniques have been widely studied in literature with the aim of reducing the congestion in the backhaul link,  the energy consumption and the latency. In cache-enabled networks content is pre-fetched close to the user during network off-peak periods in order to directly serve the users when the network is congested. In their seminal work,  Maddah-Ali \emph{et al.} \cite{Maddah2015}  aim at reducing the transmission rate in a network where each user has an individual cache memory.  In their work,  the idea of \emph{coded caching} is introduced, i.e. the cache memory does not only  provide direct local access to the content but also  generates the so called  coded multicasting opportunities among users requesting different files, making it possible to serve several users with a single transmission.

\ac{MDS}  codes have been proposed to minimize the use of the backhaul link during the delivery phase in networks with caches at the transmitter side only \cite{Caire:femto, bioglio:Globcom2015, Liao:2017}. A delayed offloading scheme based on \ac{MDS} codes was proposed to spare backhaul link resources in a network with mobile users \cite{ozfatura2018}. Caching schemes leveraging on \ac{MDS} codes have also been proposed  for device to device communication in order to reduce the latency  \cite{piemontese2016}.

Codes are classified as \emph{fixed rate} codes when their codeword blocklength  is a priory fixed. \reviewerone{\ac{MDS} codes} are optimal fixed rate codes in sense that they achieve the Singleton bound.  
The drawback of \reviewerone{MDS codes} is that the   order of the field used for their construction increases with the blocklength. 
\fran{Furthermore, when using MDS codes in a caching scheme\cite{bioglio:Globcom2015} the blocklength (and, possibly, the field order) must be chosen depending on the number of transmitters and the topology of the network. }
Unlike \ac{MDS} codes, fountain codes \cite{MacKay05:fountain}  are \emph{rateless}, i.e., their rate can be adapted on-the-fly. This has the advantage of adding flexibility to the network, allowing a dynamic resource management.

Extensive studies regarding caching for terrestrial
applications can be found in \reviewerone{the} literature. However,  the number of \reviewerone{works} considering caching in  heterogeneous satellite networks is limited. 
Caching has been studied by \reviewerone{de Cola} et al. \cite{DeCola:icn}  in satellite-assisted emergency communications to reduce end-to-end delay. 
A two-layer caching model for content delivery services in satellite-terrestrial networks has also been proposed\cite{Wu:Twolayer}. Content placement in LEO satellite constellation networks has been also proposed  in order to minimize user terminals content access delay\cite{Liu:LeoGame}. 
An  off-line caching approach over a hybrid satellite-terrestrial network  has also been  proposed  for reducing the traffic of terrestrial network\cite{Kalantari:SatTerresrial}. However, sparing backhaul resources is of particular importance not only for terrestrial networks but also in satellite systems.

In this paper we propose the application of fountain codes  for caching content  in satellite heterogeneous networks. \reviewertwo{ We consider a heterogeneous network, in which different transmitter types coexist.  In particular, we consider a satellite which acts as central entity and  has direct connectivity to the cache-enabled transmitters (hubs).  Despite the fact that a satellite architecture is assumed, our work can be adapted to terrestrial heterogeneous networks.}

We show how the performance of a fountain code based caching scheme approaches that of a scheme that uses \ac{MDS} \reviewerone{codes} in terms of backhaul transmission rate. To this end,  among the class of fountain codes, we analyze the performance of \acfp{LRFC}  which represents a benchmark for   extending in future the analysis to other types of fountain codes (i.e. LT codes, raptor codes).  In particular, we study and optimize the performance of fountain codes for caching-enabled networks with satellite backhauling. We extend our previous work\cite{ASMS:cachingLRFC} by introducing a novel and simpler derivation of the average backhaul transmission rate\footnote{We define the average backhaul transmission rate as the average number of coded packets (output symbols) that the \acs{GEO} needs to send through the backhaul link during the delivery phase to serve the request of a user.}. Furthermore, we derive a new upper bound to the average backhaul transmission rate which is tighter than the one presented in our previous work\cite{ASMS:cachingLRFC}.  Furthermore, we present additional simulation results, which show that the performance of the caching system using \er{binary} \acp{LRFC} is close to that of a system based on \ac{MDS} codes when we consider a sufficient number of input symbols. 

The rest of the paper is organized as follows. Section~\ref{sec:sysmod} introduces the system model, while in Section~\ref{sec:fountaincode} some preliminaries on \acp{LRFC} are presented. The expression of the achievable backhaul rate is presented in Section~\ref{sec:averageRate}. The optimization problem related to the number of coded symbol to be memorized at each cache is presented in  Section~\ref{sec:opt}. In Section~\ref{sec:results} the numerical results are presented. Finally, Section~\ref{sec:Conclusions} contains the conclusions. 


\section{System Model}\label{sec:sysmod}

\begin{figure}[t]
 \includegraphics[width=0.4\textwidth]{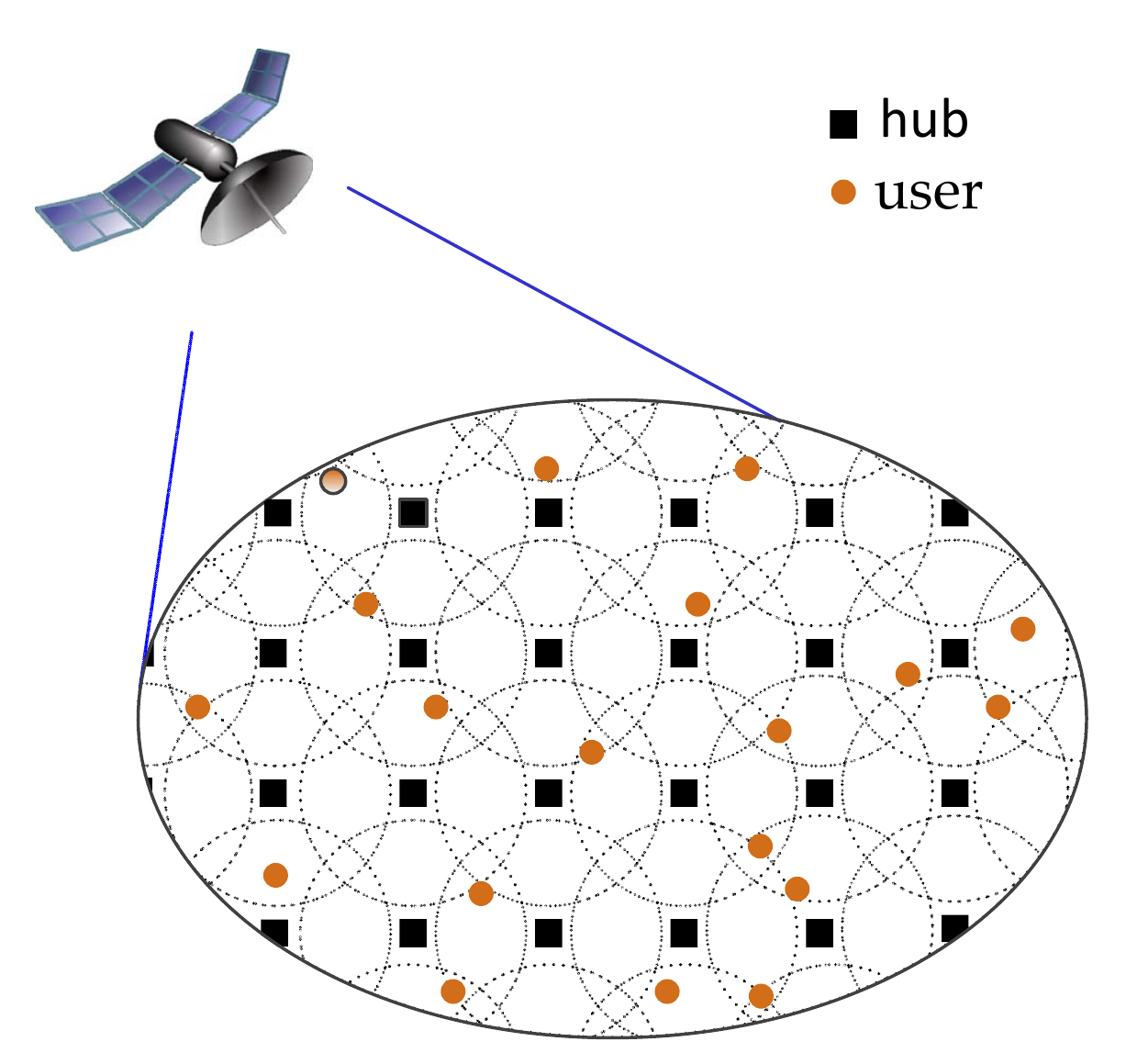}
\centering\caption{\reviewerone{System model for a satellite heterogeneous network composed of a satellite, a number of hub transmitters with caching capability and multiple users.} }
\label{fig:model}
\end{figure}

We consider a two-tier heterogeneous network composed of a
 \ac{GEO} satellite, a number of hubs (e.g. terrestrial repeaters or \reviewerone{\acp{HAP}} )  with cache capabilities and fixed users,  as shown in Fig.~\ref{fig:model}.  Each hub is connected to the \ac{GEO} satellite through a backhaul link. Users are assumed  to have a limited antenna gain so that a direct connection to the \ac{GEO} satellite is not possible. Depending on their location, users may be connected to one or multiple hubs.  We denote by $\probHub{\hubs}$ the probability that a user is connected to $\hubs$ hubs,  and we assume that $\probHub{\hubs}$  can be easily derived from the geometry of the network.

The \ac{GEO} has access to a library of $\nfiles$ files  \reviewerone{(e.g. video clips)} $\library = \{\libraryf{1}\, \ldots, \libraryf{\nfiles}\}$, all having identical size.  We assume that users request files from the library independently at random. Furthermore, we assume that the probability of file $\libraryf{j}$ being requested, $\probFile{j}$, follows a Zipf distribution\reviewerone{\cite{Zipf}} with parameter $\alpha$ leading to 
\begin{align} \label{eq:zipf}
\probFile{j} = \frac{1/j^\alpha}{\sum_{i=1}^n 1/i^{\alpha}}.  
\end{align}

We assume that the caching process in the satellite heterogeneous network consists of two phases: a placement phase and a delivery phase. In the placement phase, 
each file is  fragmented into $\npacket$ \emph{input symbols} (packets) and the \ac{GEO} satellite encodes each file independently, using a linear random fountain code. During this phase, 
\reviewertwo{the GEO satellite fills up each   cache by transmitting $\npcached{j}$  output symbols from file $\libraryf{j}$. }
Each hub has storage capability for $\memory$ files, i.e., for $ M k$ packets, such that the  following  holds for every cache \[ 
\sum_{j=1}^n \npcached{j} = \memory  k.
\]
We want to highlight that in our scheme  each hub caches for each file the same number 
of \emph{output symbols} (encoded packets). However, 
different   sets of output symbols are cached at different hubs. 
We also remark that the placement phase is  carried out offline.
In the delivery phase, users request files  at random. In a first stage, the user downloads $m = w_j \cdot h$ different output symbols of $\libraryf{j}$ cached in the $\hubs$ hubs he is connected to. 
 Whenever the number of symbols received is not enough for decoding  $\libraryf{j}$ successfully, additional output symbols must be sent through the backhaul link to one of the neighbouring hubs, which forwards them to the users. 
For simplicity we assume that all transmissions are error-free.

\section{Linear Random Fountain Codes}
\label{sec:fountaincode}

In this work we consider the use of \acp{LRFC} for the delivery of the different files in the library. Each file is  fragmented into $k$ input   symbols,
$ (\isymb_1, \isymb_2, \hdots, \isymb_\npacket )$. 
\reviewerone{For simplicity, we assume that the input symbols $\isymb_i$ belong to a finite field of order $q$ i.e., $\isymb_i \in \mathbb{F}_q$}. The case of  $\isymb_i \in \mathbb{F}_q^m$, i.e. the case in which packets are $m$ symbols long,  can be addressed as a straightforward extension. The \ac{LRFC} encoder generates a sequence of output symbols
$\mathbf{\codedpacket} = ( \codedpacket_1, \codedpacket_2, \hdots, \codedpacket_{\ell} )$,
where the number of outputs symbols $\ell$ can grow indefinitely. In particular, the $i$-th output symbols is generated as
\[
\codedpacket_{\indexi} = \sum_{a=1}^{\npacket} \coeff_{a,\indexi}   \, \isymb_{a},
\]
where the coefficients $\coeff_{a,\indexi}$ are picked  independently at random with uniform probability in $\mathbb{F}_q$.
For fixed $\ell$, \ac{LRFC} encoding can be expressed as a vector matrix multiplication
\[
\mathbf{\codedpacket} = \isymbv \G
\]
where $\isymbv$ is the vector of input symbols and $\G$ is a $k \times \ell$ matrix with entry $\coeff_{a,\indexi}$. In order to download a file, a user must collect a set of $m \geq \npacket$ output symbols $\rosymbv=(\rosymb_1, \rosymb_2, \hdots, \rosymb_m)$. If we denote by ${\mathcal{I} =(i_1, i_2, \hdots, i_m)}$ the set of indices corresponding to the $m$ output symbols collected by the receiver we have
\[
\rosymb_r = \codedpacket_{i_r}.
\]
The user attempts decoding by solving the system of equations
\[
\rosymbv = \isymbv \Grx
\]
where $\Grx$ is a matrix corresponding to the $m$ columns of $\G$ associated to the collected output symbols i.e., the columns of $\G$ with indices in $\mathcal{I}$. If the system of equations  admits a unique solution (i.e., if $\Grx$ is full rank), decoding is declared successful after recovering $\isymbv$, for example by means of Gaussian elimination. If $\Grx$ is rank deficient, a decoding failure is declared. In the latter case the receiver reattempts decoding after collecting one or more additional output symbols.

Let us define $\delta$ as the receiver overhead  $\delta=m-\npacket$, that is, the number of output symbols in excess to $\npacket$ that the receiver has collected.
Given $k$, $\delta$ and $q$,  the probability of decoding failure of an \ac{LRFC} is given by 
\[
\probFail{(\npacket,\delta,q)} = 1- \prod_{\indexi=1}^{\npacket} \left(  1-\frac{q^{\indexi-1}}{q^{\npacket+\delta}}  \right)
\]
and can be tightly lower and upper bounded as  \cite{Liva10:fountain}
\begin{align}\label{eq:bounds}
\lowerb{(\delta,q)} & \leq \probFail{(\npacket,\delta,q)}  <   \upperb{(\delta,q)}
\end{align}
where
\[
\lowerb{(\delta,q)}: = q^{-\delta -1}
\]
and
\[
 \upperb{(\delta,q)}: =  \frac{1}{q-1} q^{-\delta}.
\]
Note that the bounds are independent from the number of  input symbols $\npacket$ and become tighter for increasing $q$.

 For notational convenience, in the remaining of the paper we 
denote    $\probFail{(\npacket,\delta,q)}$ by $P_F(\delta)$, that is, we do not indicate explicitly the dependency on $\npacket$ and $q$.


\section{Average Backhaul Transmission Rate}\label{sec:averageRate}

We define the average backhaul transmission rate $  \mathbb{E}[T]$ as the average  number output coded symbols that the \ac{GEO} satellite has to transmit during the delivery phase in order to fulfill  a user  request.

\subsection{Overhead Decoding Probability} \label{susec:DecProb}
We define the overhead decoding probability $\Psuc{\delta}$ as the probability that a user needs \emph{exactly} $k+\delta$ coded  symbols to successfully decode the requested file.  
Hence,  $\Psuc{\delta}$ corresponds to the probability that the matrix $\Grx$ is full rank  when $m = k + \delta$ output symbols have been collected, \fran{conditioned to the fact that matrix $\Grx$ was not full rank when $m=k+ \delta -1$ symbols had been collected}. For a fixed number of input symbols $k$ and a fixed field order $q$,  $\Psuc{\delta}$ can be written as
\[ \Psuc{\delta} = P_F(\delta-1)-P_F(\delta).
\]

\subsection{Overhead Average}
 Let us denote as  $\Delta$ the random variable associated   to the  average  number of symbols in excess to $k$  that a user needs in order to recover the requested content and  let us also denote as $\delta$ its realization. We can calculate  the average overhead as follows
\begin{align}
\Expd{} & = \sum_{\delta=1}^{\infty} \delta  \cdot \big[ P_F(\delta-1)-P_F(\delta) \big] \nonumber\\
 & = \sum_{\delta=0}^{\infty} (\delta+1) \cdot P_F(\delta) -  \sum_{\delta=0}^{\infty} \delta \cdot P_F(\delta) \nonumber  \\
  & = \sum_{\delta=0}^{\infty}  P_F(\delta).\label{eq:delta_av}
\end{align}

By using \eqref{eq:bounds} in \eqref{eq:delta_av}, $\Expd{}$ can be upper bounded as
\begin{align}
\Expd{}&  <  \sum_{\delta=0}^{\infty} \frac{q^{-\delta}}{q-1} \nonumber \\
& \stackrel{(\mathrm{a})}{=}  \frac{q}{(q-1)^2} :=  \delta_u.   \label{eq:delta_u}
\end{align}
 

\subsection{Backhaul Rate}
\fran{Let us now consider a generic user requesting a file.
Let $Z$ be the random variable associated to the number of output symbols from the requested file which are available at the hubs that the user is connected to. Also let $z$ be the realization of $Z$.  Let $H$ be the random variable associated to the number of hubs a user is connected to, being $h$ its realization. Finally, let $J$ be the random variable associated to the index of the file requested by the user, being $j$ its realization.} We have
\begin{align}\label{eq:Z}
  P_{Z|J,H}(z | j, h) = \begin{cases}
                              1  & \mbox{if } z= \npcached{j} \, h\\
                              0  & \mbox{otherwise}
                            \end{cases}
\end{align}
where we recall that  $\npcached{j}$ stands for the number of coded symbols from file $j$ stored in every hub. The probability mass function of $Z$ is
\begin{align}\label{eq:z_2}
    P_Z(z)    &= \sum_j \sum_h  P_{Z|J,H}(z | j, h) P_J(j)  P_H(h) \nonumber  \\
   & =  \sum_j \sum_h \probFile{j} \, \probHub{\hubs} \,  P_{Z|J,H}(z | j, h).
\end{align}
We are interested in deriving the average backhaul transmission rate, i.e, the average number of output symbols which have to be sent over the backhaul link to serve the request of a user, which we denote by random variable $T$. \er{The conditional probability of $T$ given $Z$ corresponds to the   decoding success probability when exactly $t$ output symbols have been received from the backhaul link. A user collects in total $m=z+t$ output symbols where $z$ symbols are transmitted from the caches through local links.  } In order to derive $P_{T|Z}(t|z)$ we shall distinguish two cases.
\newline
 If $z \leq k$, then
\begin{align}\label{eq:t_cases1}
&  P_{T|Z}(t|z) =
\begin{cases}
 \displaystyle    \Pfail(z-k+t-1) - \Pfail(z-k+t ) \mkern15mu   \mbox{if } t \geq 0   \\
  0  \mkern255mu  \mbox{otherwise.}
\end{cases}
\end{align}
\newline
If $z>k$, we have
\begin{align}\label{eq:t_cases2}
&  P_{T|Z}(t|z) =
\begin{cases}
 \displaystyle  1 - \Pfail(z-k)   \mkern165mu   \mbox{if } t = 0,& \\  \displaystyle  \Pfail(z-k+t-1) - \Pfail(z-k+t )  \mkern18mu \mbox{if t} > 0,& \\
                   0    \mkern258mu  \mbox{otherwise.} &
  \end{cases}
\end{align}
\newline
The expectation $\mathbb{E}[T]$ is given by

\begin{align}\label{eq:t}
  \mathbb{E}[T] = &\sum_t t \Bigg( \sum_{z=0}^{k} \Pro_{T|Z}(t|z) \Pro_Z(z)   + \sum_{z=k+1}^{\infty} \Pro_{T|Z}(t|z) \Pro_Z(z)  \Bigg).   
\end{align}
\newline
\reviewerone{As shown in the Appendix, it is possible to upper bound \eqref{eq:t} as}
\begin{equation}
\mathbb{E}[T] \leq 
  \delta_u +    \sum_{z=0}^{k}(k- z)\Pro_Z(z). 
\label{eq:ETbound}
\end{equation}


\section{LRFC Placement Optimization Problem}\label{sec:opt}
The \ac{LRFC} placement   problem calls for
minimizing the average backhaul transmission rate during the delivery phase. In particular, we would like to determine the number of coded symbols per file that each hub has to cache, so that the average backhaul transmission rate is minimized.  We present in this section the placement optimization problem adapted to a \ac{LRFC} cached scheme based on the optimization problem proposed for \ac{MDS} codes in~\cite{bioglio:Globcom2015}.

The optimization problem can be written as
\begin{align} \label{eq:optProblem}
\underset{\npcached{1}, \dots, \npcached{\nfiles}}  \min &  \quad  \mathbb{E}[T]  \\
 \text{\reviewerone{subject to} }& \quad \sum_{j=1}^{\nfiles}  \npcached{j}    = \memory \npacket \nonumber\\
 & \quad     \npcached{j} \in  \mathbb{N}_0. \nonumber
\end{align}
The first constraint specifies that the total number of stored coded symbols should be equal to the size cache. The second constraints accounts for the discrete nature of the optimization variable.

Solving exactly the optimization problem requires evaluating \eqref{eq:t},  which is complex. Hence, as an alternative to minimizing the average backhaul  transmission rate, we propose minimizing its upper bound  in~\eqref{eq:ETbound}, which leads to the following optimization problem
\begin{align}\label{eq:optProblem2}
\underset{\npcached{1}, \dots, \npcached{\nfiles}}  \min & \quad \Bigg[   \delta_u   +  \sum_{z=0}^{k} (k-z)  \Pro_Z(z)  \Bigg] \\
 \text{\reviewerone{subject to} }& \quad \sum_{j=1}^{\nfiles}  \npcached{j}    = \memory \npacket  \nonumber\\& \quad  \npcached{j} \in \mathbb{N}_0.  \nonumber
\end{align}
Since the upper bound on $\mathbb{E}[T]$  in \eqref{eq:ETbound} relies on the upper bound in \eqref{eq:bounds}, which is tight, we expect the result of the optimization problem in \eqref{eq:optProblem2} to be close to the result of the optimization problem in \eqref{eq:optProblem}. 


\section{Results}\label{sec:results}
In this section, we numerically evaluate the normalized average backhaul transmission rate, which we define as \[ \widehat{T} = \frac{\mathbb{E}[T]}{k}.\]

In all the setups, we consider that users are uniformly distributed within the coverage area of the satellite and border effects are neglected.
 We consider  that each hub covers a circular area of radius $\radius{}$ centered around the hub.   For simplicity, we assume that the hubs are arranged according to a uniform two dimensional grid, with spacing $\dcenter$.  Unless otherwise specified, we assume $\radius{}=60$ km and   $\dcenter=45$ km. Thus, the coverage areas of different hubs partially overlap, as it can be observed in Fig.~\ref{fig:model}. With geometrical calculations  the following connectivity distribution can be obtained  
\begin{align} \label{eq:gammadis}
& \gamma_1 = 0.2907,  \,  \gamma_2 =0.6591,   \\
& \gamma_3 = 0.0430, \, \gamma_4 =0.0072.  \nonumber\
\end{align}

We first evaluate the tightness of the upper bound \eqref{eq:delta_u} on the average overhead. Table~\ref{tab:tabdelta} shows  $\Expd{}$  for different values of $q$. The values in the second column were numerically derived from equation~\eqref{eq:delta_av} while values in the third column were derived from the bound in equation~\eqref{eq:delta_u}. We can see that the bound becomes tighter for increasing $q$.

\begin{center}
\begin{table}[t] \centering
\caption{Average overhead required for successful decoding for a \ac{LRFC} }\label{tab:tabdelta}
\begin{tabular}{ccc}
\hline
  $q$   &  $\Expd{}$   &  $\delta_u$ \\
\hline
$2$   & 1.6047     &     2  \\
$4 $  & 0.4211     & 0.4444 \\
$8 $  & 0.1610     & 0.1633 \\
$16$  & 0.0708     & 0.0711 \\
$32$  & 0.0333    & 0.0333 \\
$64$  & 0.0161    & 0.0161 \\
$128$ & 0.0079     & 0.0079 \\
\hline
\end{tabular}
\end{table}
\end{center}
 
In the first scenario, we study the impact of the cache size $M$ on the average backhaul transmission rate.  In this setup,
we consider the connectivity distribution  given in~(\ref{eq:gammadis}) and file popularity distribution given by \eqref{eq:zipf} with parameter $\alpha =0.8$. The library size is set to $n=100$. 
We optimized  the number of LRFC coded symbols $\npcached{j}$  cached  at each hub by solving the problem \eqref{eq:optProblem2} for  $q=2$, $q=4$ and $q=16$, and we  calculated numerically the average backhaul transmission rate of our fountain coding caching scheme by applying~\eqref{eq:ET}. As a benchmark, we used the \ac{MDS} caching scheme from Bioglio et al.~\cite{bioglio:Globcom2015}.  \reviewertwo{We would also like to remark that the performance of a scheme without  caching is characterized in our setting by   $\widehat{T}=1$, since all the content has to be transmitted through the backhaul link.}

\begin{figure}[t]
 \includegraphics[width=0.55\textwidth]{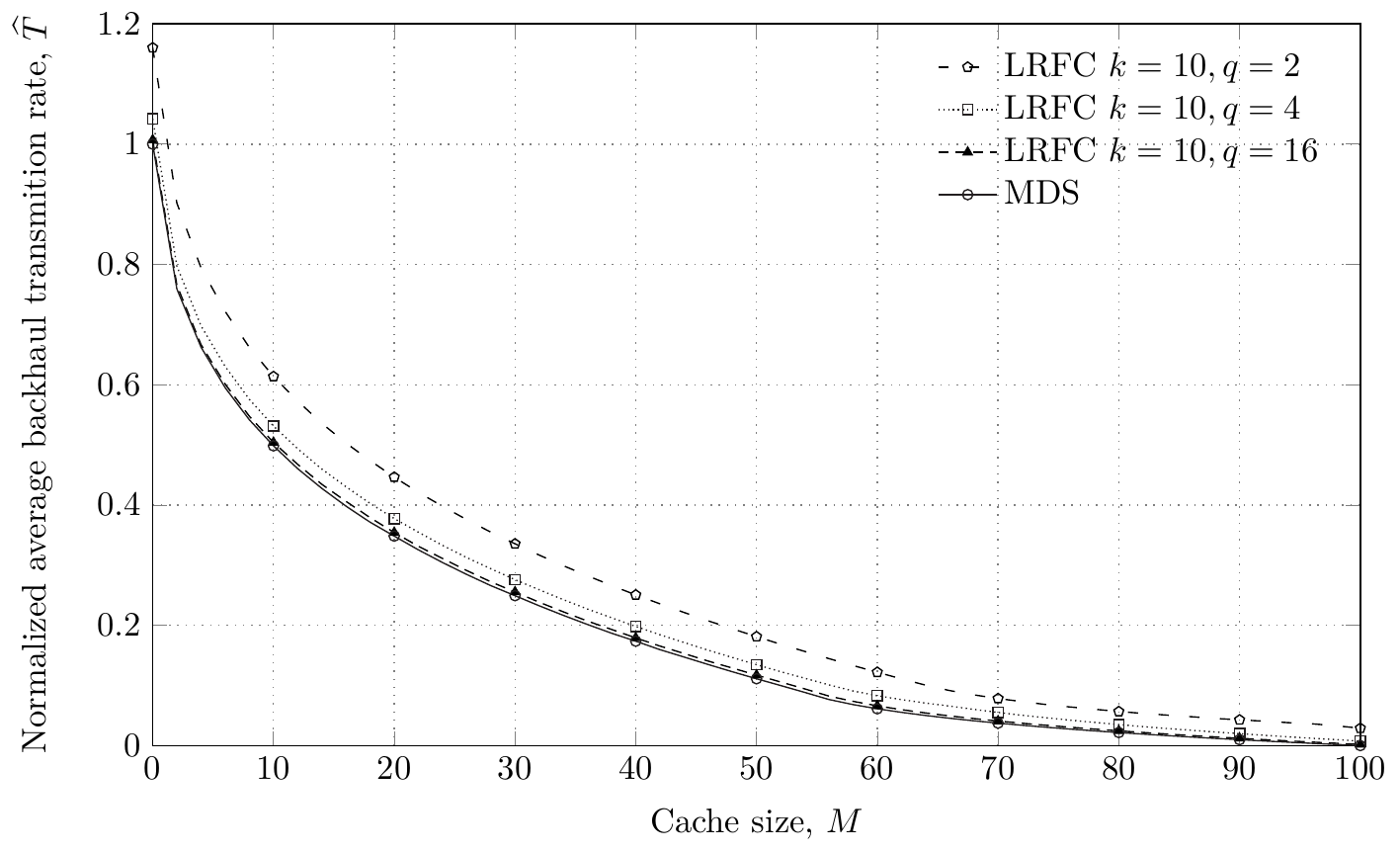}
\centering \caption{Normalized average backhaul transmission rate as a function of memory size $M$ for  \er{ \ac{LRFC} codes over $\mathbb{F}_q$ for ${q=2,4,16}$ with $k=10$  and for an \ac{MDS} code,} given $n=100$,  $\alpha=0.8$ and  $\gamma_1=0.2907$, $\gamma_2=0.6591$, $\gamma_3=0.0430$, $\gamma_4=0.0072$. Solid curves represent \ac{LRFC} schemes while the dashed curve represents the \ac{MDS} scheme.}
\label{fig:R_m10}
\end{figure}

In Fig.~\ref{fig:R_m10} the normalized average backhaul transmission rate is shown as a function of the memory size $M$ when each file is fragmented into $k=10$ input symbols. We can observe how the penalty on the average  rate for using \ac{LRFC} with respect to a \ac{MDS} code becomes smaller for increasing $q$ and for $q = 16$ is almost negligible. We remark that for $M=100$ the cache size coincides with the library size, hence, the backhaul  rate for the \ac{MDS} scheme becomes zero, whereas for the \ac{LRFC} schemes the average backhaul  transmission rate coincides with the average overhead.
 Note that since a MDS code achieves the best possible performance, this result shows implicitly that solving the optimization problem in \eqref{eq:optProblem2} yields a solution that is close to that of solving the  optimization problem in \eqref{eq:optProblem}. We further observe that  \ac{LRFC}  caching     with storage capabilities  equal  to 10\% of the library size can reduce the average backhaul  rate for at least 40\%  with respect to a system with no caching ($M=0$).

\begin{figure}[t]
 \includegraphics[width=0.55\textwidth]{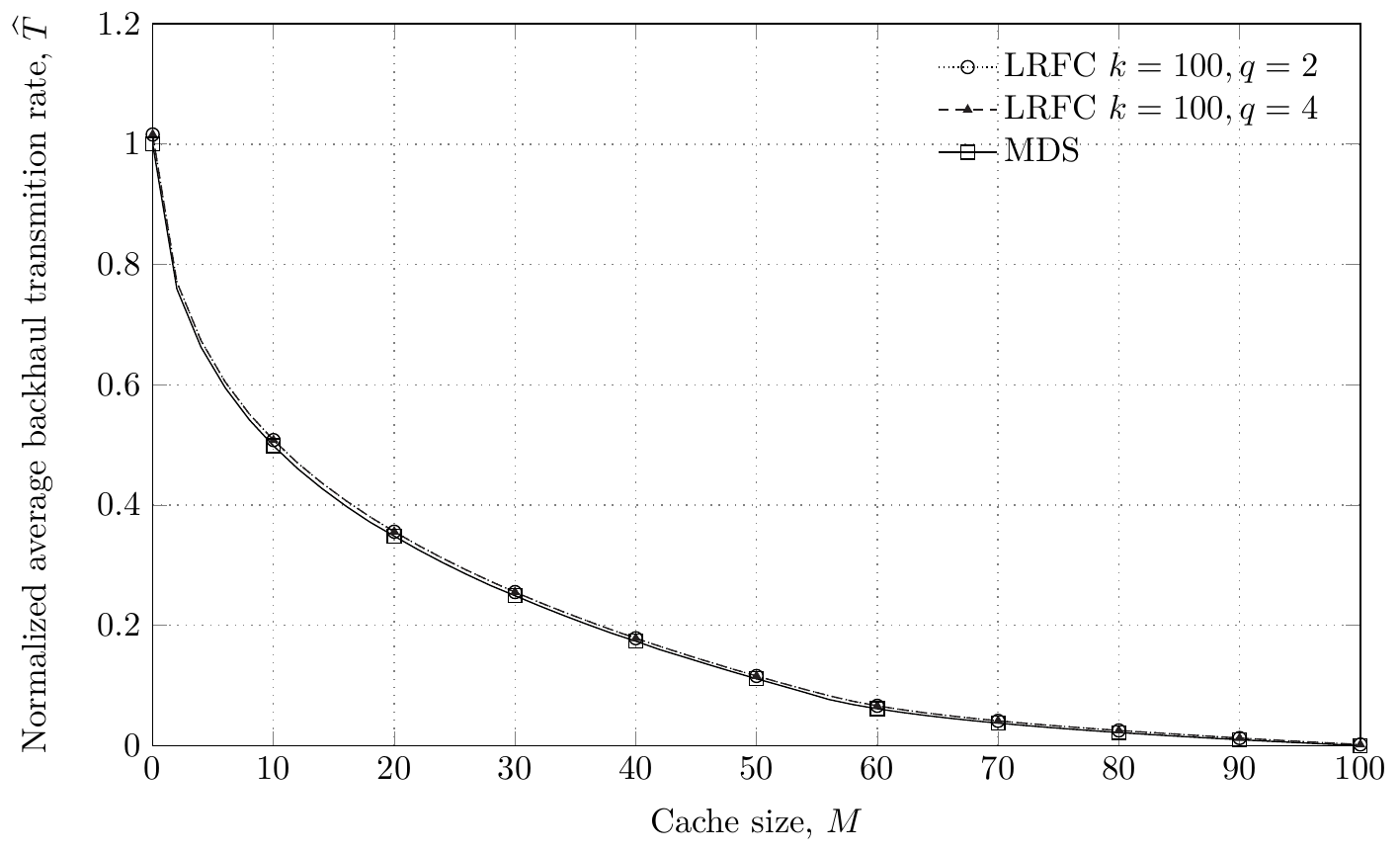}
\centering \caption{Normalized average backhaul transmission rate as a function of memory size $M$ for \er{ \ac{LRFC} codes over $\mathbb{F}_q$ for $q=2,4$ with $k=100$  for and for an \ac{MDS} code,}  given $n=100$,  $\alpha=0.8$ and $\gamma_1=0.2907$, $\gamma_2=0.6591$, $\gamma_3=0.0430$, $\gamma_4=0.0072$.}
\label{fig:R_M100}
\end{figure}

Next, we consider the same parameters as the previous scenario but we evaluate the average backhaul transmission rate when each file is fragmented into $k=100$ input symbols.  In Fig.~\ref{fig:R_M100} we show the impact of memory size $M$  on the normalized average backhaul transmission rate  for different code caching schemes when $k=100$. We can observe that the 
 optimal cache placing  computed for  an \ac{LRFC} caching scheme built in a field of order $q=2$ reaches the performance of the \ac{MDS} scheme and slightly better performance is given when $q=4$. Thus,  the \ac{LRFC} caching scheme can  approach the best performance in the binary field  just by increasing the number of input symbols. 
 
\begin{figure}[t]
 \includegraphics[width=0.55\textwidth]{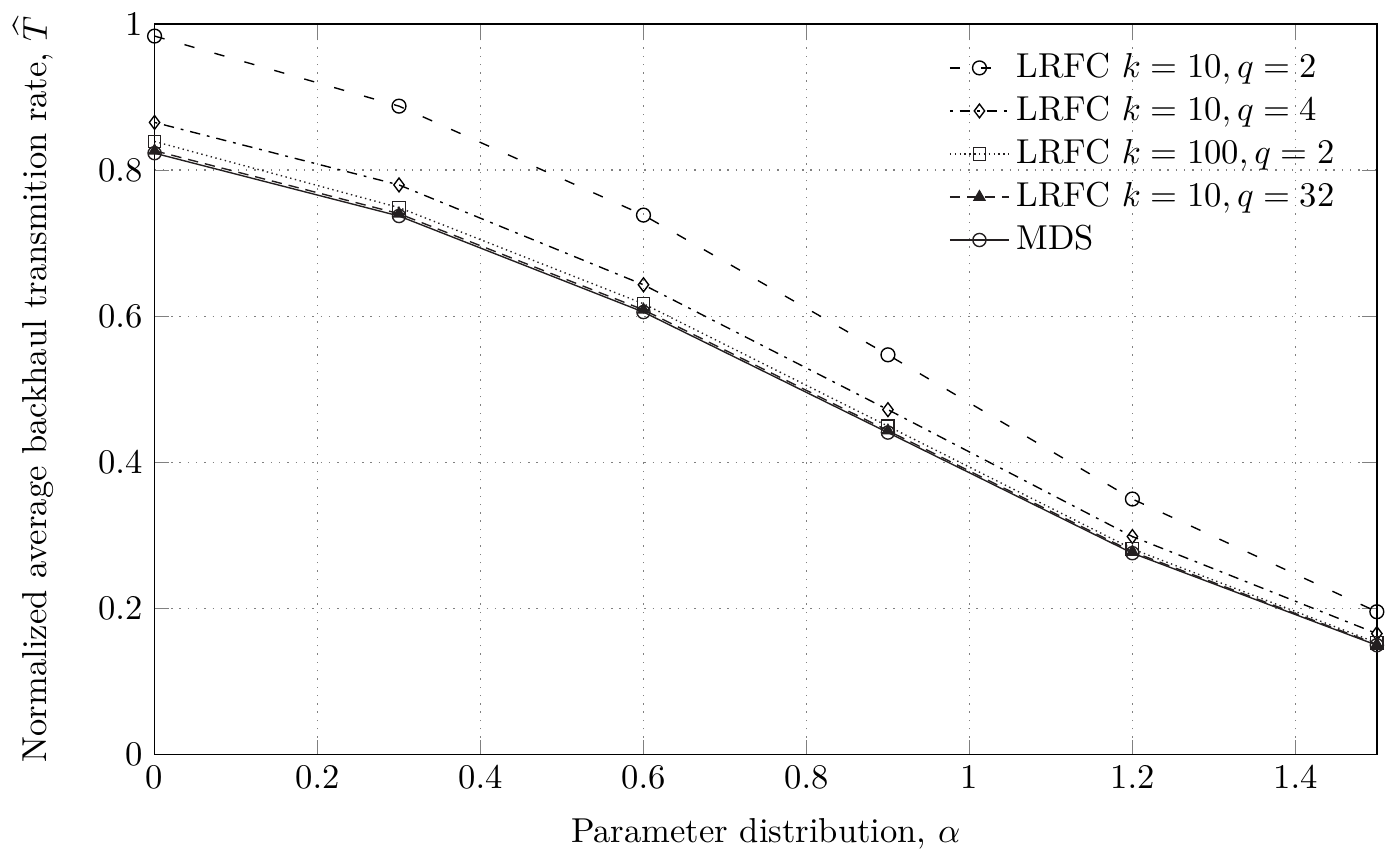}
\centering \caption{Normalized average backhaul transmission rate as a function of the file parameter distribution $\alpha$ for \er{ \ac{LRFC} codes over $\mathbb{F}_q$ for $q=2,4,32$ with $k=10$ or $k=100$ and for an \ac{MDS} code,}  given $n=100$,  $M=10$  $\gamma_1=0.2907$, $\gamma_2=0.6591$, $\gamma_3=0.0430$, $\gamma_4=0.0072$.}
\label{fig:R_alpha}
\end{figure}

For the same connectivity distribution, library size $n=100$ and fixed memory at each cache of $M=10$, we  investigate how the file distribution impacts on the average backhaul transmission  rate.
In Fig.~\ref{fig:R_alpha} the normalized average backhaul  transmission rate is shown as a function of the file parameter distribution $\alpha$. As expected, when $\alpha$ increases, caching schemes become more efficient since the majority of the requests is concentrated in a small number of files.
Looking at the figure we can observe how for $\alpha=0.2$, a \ac{LRFC} in  $\mathbb{F}_2$ with $k=10$ requires roughly 12\% more transmissions over backhaul link   than  a  \ac{LRFC} in  $\mathbb{F}_{32}$ with $k=10$. We can also observe that a binary \ac{LRFC}   with $k=100$ performs as good as the optimal scheme.  
 For  $\alpha=1.5$ the \ac{LRFC} of order $q=2$ with $k=10$ requires only 4.7\% more than in $q=2$ with $k=100$.

\begin{figure}[t]
 \includegraphics[width=0.55\textwidth]{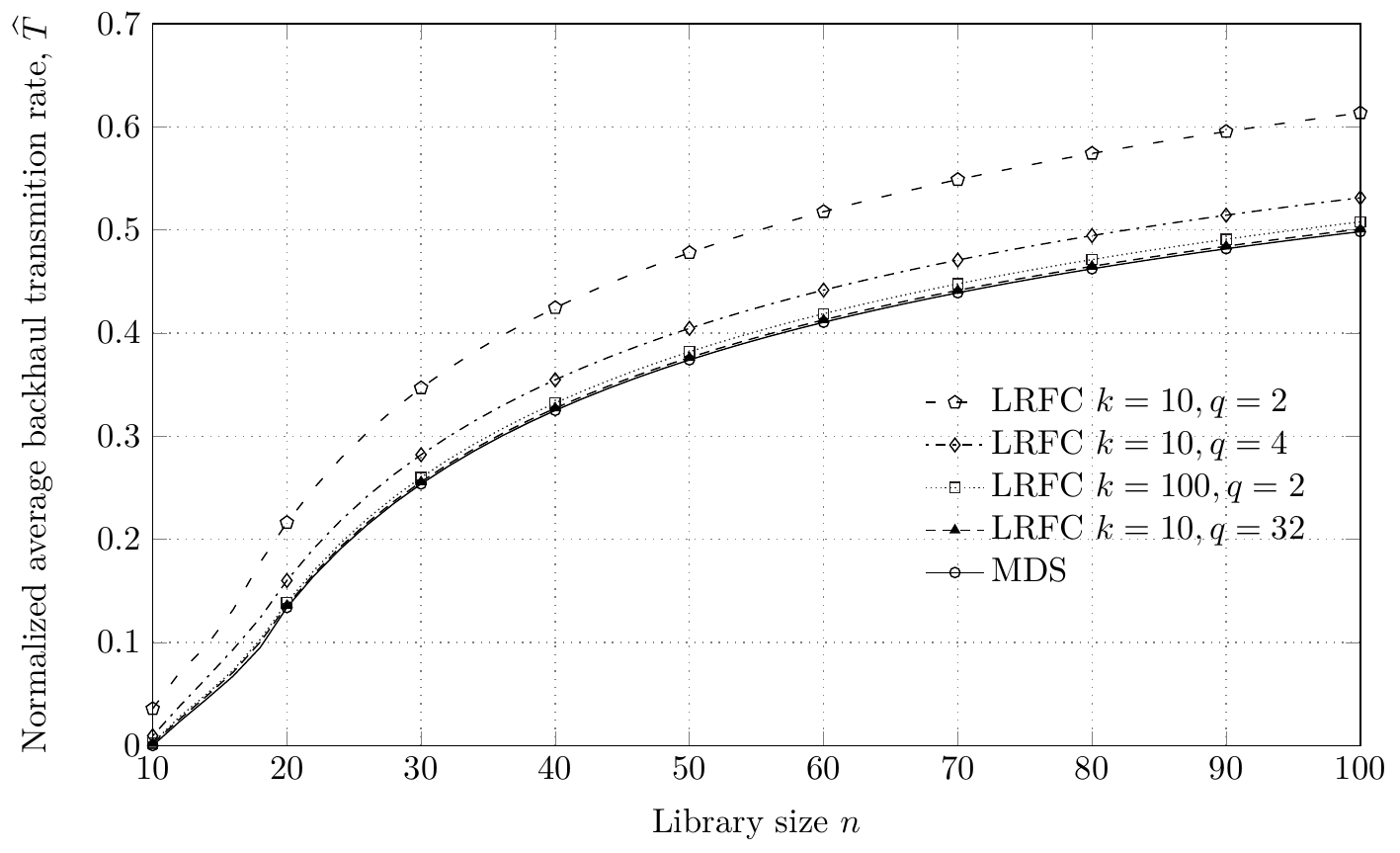}
\centering \caption{Normalized average backhaul transmission rate as a function of the library size $n$ for  \er{ \ac{LRFC} codes over $\mathbb{F}_q$ for $q=2,4,32$ with  $k=10$  and for an \ac{MDS} code,} given $\alpha=0.8$, $M=10$ and $\gamma_1=0.2907$, $\gamma_2=0.6591$, $\gamma_3=0.0430$, $\gamma_4=0.0072$.}
\label{fig:R_n}
\end{figure}

In our next setup we consider $\alpha=0.8$, $M=10$, $k=10$ or $k=100$ and the distribution given in~(\ref{eq:gammadis}). We evaluate the average backhaul  transmission rate   for different cardinalities of the library. 
In Fig.~\ref{fig:R_n} the normalized average backhaul transmission rate is shown as a function of the library size.
For a fixed memory size the average backhaul transmission rate increases as the library size increases.  As it can be observed, also in this case the proposed \ac{LRFC} caching scheme  performs similarly to a \ac{MDS} scheme.

\begin{figure}[t]
 \includegraphics[width=0.55\textwidth]{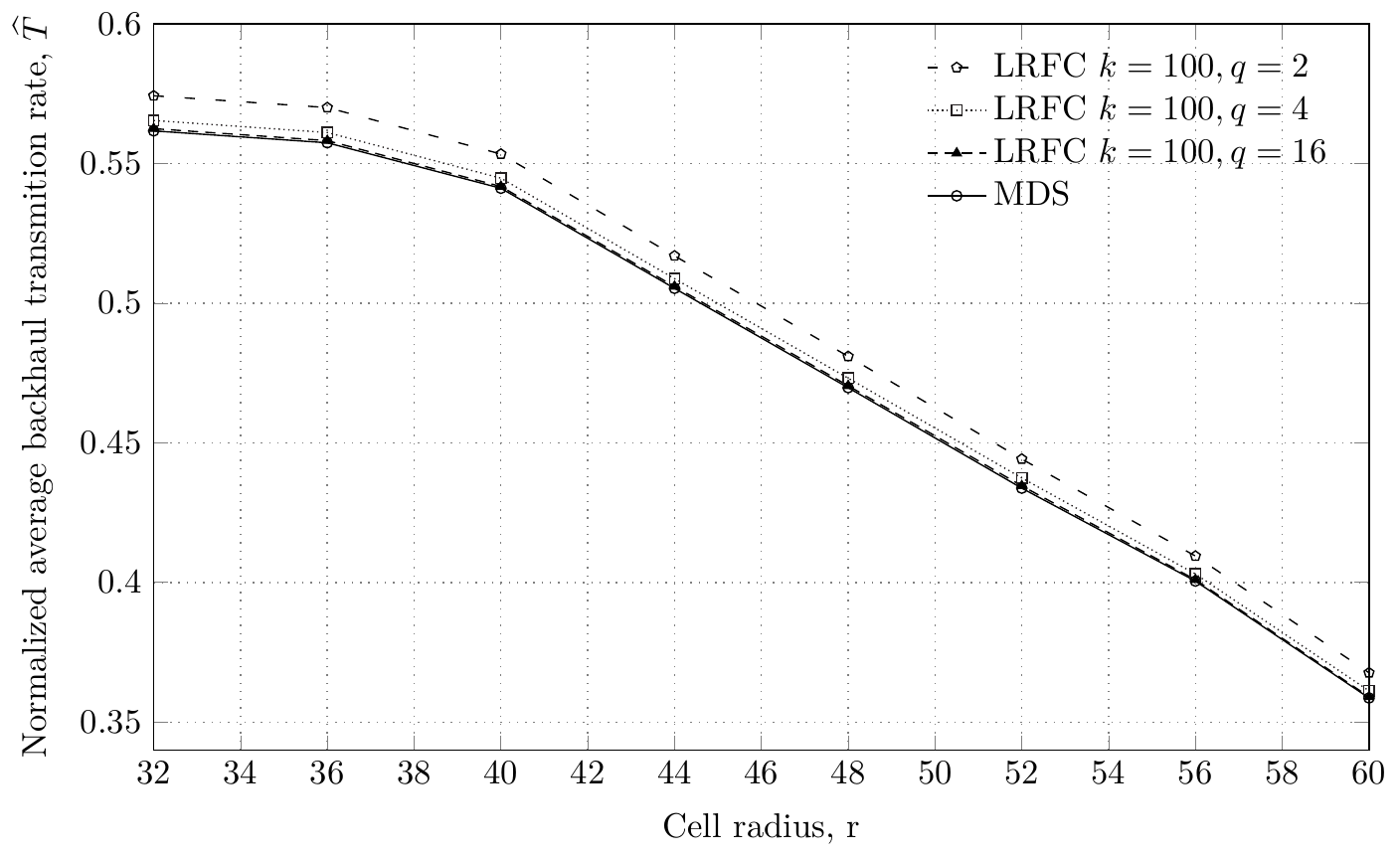}
\centering \caption{Normalized average backhaul transmission rate as a function of coverage radius $r$ for \er{ \ac{LRFC} codes over $\mathbb{F}_q$ for $q=2,4,16$ with $k=100$ and for an \ac{MDS} code,} given $\alpha=0.8$ and  $M=10$.}
\label{fig:R_r}
\end{figure}

In our last setup we consider  $M=10$, $k=100$   and   $\alpha =0.8$. We compute the average backhaul transmission rate in function of the radius of coverage $r$ of the hubs. The plot in Figure~\ref{fig:R_r} shows that gain of using a \ac{MDS} caching scheme with respect of the binary \ac{LRFC} caching scheme is at maximum of the 0.0126\% when the hubs has radius of coverage $r=32$ km and  becomes smaller for larger values.

\newpage
\section{Conclusions}\label{sec:Conclusions}
\reviewertwo{In this work, we analyze a caching scheme based on fountain codes, which are characterized by a high degree of flexibility.
In particular, we derived the analytical expression of the average backhaul transmission rate and a simple yet tight upper bound to it.  The optimization problem related to which content to place in the cache memories was formalized and solved. Numerical results were presented comparing the fountain code caching scheme with a caching scheme based on MDS codes available in literature. The results indicate that the fountain code caching scheme approaches the performance of the MDS caching scheme when the number of input symbols is high, or when the fountain code is built over finite field of moderate high order.}

 \newpage
 \reviewerone{\section*{Appendix}\label{sec:Appendix} }
Let us define $\bar{T}_1$ and $\bar{T}_2$ as
\begin{align}
  \bar{T}_1 :&=  \sum_t t  \sum_{z=0}^{k}\Pro_{T|Z}(t|z) \Pro_Z(z) \nonumber\\
  \bar{T}_2 :&=  \sum_t t \sum_{z=k+1}^{\infty} \Pro_{T|Z}(t|z)  \Pro_Z(z) \nonumber
\end{align}
so that
\begin{align} \label{eq:T1T2}
 \mathbb{E}[T] = \bar{T}_1 + \bar{T}_2.
\end{align}
\newline
If we introduce the variable change $\delta = z-k+t$ in the expression of $\bar{T}_1$, we obtain
\begin{align}
  \bar{T}_1 & =
  \sum_{z=0}^{k}  \Pro_Z(z)  \sum_{\delta =z-k}^{\infty} (\delta  -z+k )\Big[\Pfail(\delta - 1)-\Pfail(\delta )\Big] \nonumber  \\
  & \stackrel{(\mathrm{a})}{=}  \sum_{z=0}^{k} \Pro_Z(z)  \sum_{\delta  = 0}^{\infty} (\delta  -z+k )\Big[\Pfail(\delta- 1)-\Pfail(\delta )\Big]  \nonumber  \\
  &=    \sum_{z=0}^{k}  \Pro_Z(z) \Bigg( \sum_{\delta=0}^{\infty} \delta \Big[\Pfail(\delta - 1)-\Pfail(\delta )\Big]    +  (k-z) \sum_{\delta=0}^{\infty}  \Big[\Pfail(\delta - 1)-\Pfail(\delta )\Big]  \Bigg)  \nonumber  \\
  &\stackrel{(\mathrm{b})}{=}    \sum_{z=0}^{k} \Pro_Z(z)   \Big(  \Expd{} +  k-z \Big)   \nonumber  \\
  &= ( \Expd{} +  k)  \Pr\{Z \leq k\}   -  \sum_{z=0}^{k} z   \Pro_Z(z)  \label{eq:T1}
\end{align}
where equality $(\mathrm{a})$ is due to $[\Pfail(\delta - 1)-\Pfail(\delta)]=0$ for $\delta < 0$, and equality $(\mathrm{b})$ is due to
\[
\sum_{\delta=0}^{\infty} [\Pfail(\delta - 1)-\Pfail(\delta )]  = 1.
\]
\newline
If we introduce the same variable change in the expression of $\bar{T}_2$ we have
\begin{align}
  \bar{T}_2 &=   \sum_{z=k+1}^{\infty}  \Pro_Z(z)  \sum_{\delta=z-k}^{\infty} (\delta -z+k )\Big[\Pfail(\delta - 1)-\Pfail(\delta )\Big]  \nonumber  \\
  & = \sum_{z=k+1}^{\infty}  \Pro_Z(z)  \Bigg( \sum_{\delta=z-k}^{\infty} \delta \Big[\Pfail(\delta - 1)-\Pfail(\delta )\Big]   + \sum_{\delta=z-k}^{\infty} (k-z) \Big[\Pfail(\delta - 1)-\Pfail(\delta )\Big] \Bigg).\label{eq:upT}
\end{align}
Let us rewrite   \eqref{eq:upT} as follow
\begin{align}
  \bar{T}_2 &=   \sum_{z=k+1}^{\infty} \Pro_Z(z) \Big(\bar{T}_{21}(z) +  \bar{T}_{22}(z)\Big)   \label{eq:T2}
 \end{align}
 where
\begin{align}
  \bar{T}_{21}(z) &=
  \sum_{\delta=z-k}^{\infty} \delta  \Big[\Pfail(\delta - 1)-\Pfail(\delta )\Big] \nonumber\\
&= \sum_{\delta=0}^{\infty} \delta \Big[\Pfail(\delta - 1)-\Pfail(\delta )\Big] - \sum_{\delta=0}^{z-k-1} \delta \Big[\Pfail(\delta - 1)-\Pfail(\delta )\Big]  \nonumber\\
&= \sum_{\delta= 0}^{\infty} \Pfail(\delta)  - \Bigg[ \sum_{\delta=0}^{z-k-1}  \Pfail(\delta) -(z-k)\Pfail(z-k-1)  \Bigg] \nonumber\\
&= \Expd{} - \sum_{\delta=0}^{z-k-1}  \Pfail(\delta) +(z-k) \Pfail(z-k-1) \label{eq:T21}
\end{align}
and
\begin{align}
\bar{T}_{22}(z)
&= \sum_{\delta=z-k}^{\infty} (k-z) \Big[\Pfail(\delta - 1)-\Pfail(\delta )\Big]  \nonumber\\
&= (k-z) \Bigg\{ \sum_{\delta=0}^{\infty}  \Big[\Pfail(\delta - 1)-\Pfail(\delta )\Big] 
- \sum_{\delta=0}^{z-k-1}  \Big[\Pfail(\delta - 1)-\Pfail(\delta )\Big] \Bigg\} \nonumber \\
&=  (k-z) \Big\{1 - \Big[1 -\Pfail(z-k-1)\Big] \Big\}  \nonumber \\
&=  (k-z)  \Pfail(z-k-1)     \label{eq:T22}.
\end{align}
By inserting \eqref{eq:T21} and  \eqref{eq:T22} in \eqref{eq:T2} and sum we obtain
\begin{align}
\bar{T}_2 &=   \sum_{z=k+1}^{\infty}  \Pro_Z(z) \Big[ \Expd{}- \sum_{\delta=0}^{z-k-1}  \Pfail(\delta)   \Big ]\label{eq:T2f}  \\
& \leq  \sum_{z=k+1}^{\infty}  \Pro_Z(z) \Expd{} \label{eq:T2b}
\end{align}
where the inequality is due to \[\sum_{\delta=0}^{z-k-1}  \Pfail(\delta) \geq 0. \]
If we replace \eqref{eq:T1} and \eqref{eq:T2f} in \eqref{eq:T1T2}, the expression of the average backhaul transmission rate becomes
\begin{align} \label{eq:ET}
\mathbb{E}[T] &  =
   \Big( \Expd{} +  k \Big)  \Pr\{Z \leq k\}   -  \sum_{z=0}^{k} z   \Pro_Z(z)      +  \sum_{z=k+1}^{\infty}\Pro_Z(z) \Big[ \Expd{}- \sum_{\delta=0}^{z-k-1}  \Pfail(\delta)  \Big]  \nonumber\\
 & =  \Expd{} +    \sum_{z=0}^{k}(k- z)\Pro_Z(z)
  - \sum_{z=k+1}^{\infty}  \Pro_Z(z) \Bigg[  \sum_{\delta=0}^{z-k-1}  \Pfail(\delta)\Bigg]. 
\end{align}\label{eq:ETu}

Finally, we can upper bound the average transmission rate by use of \eqref{eq:delta_u} and  \eqref{eq:T2b} in \eqref{eq:ET} as 
\begin{equation}
\mathbb{E}[T] \leq 
  \delta_u +    \sum_{z=0}^{k}(k- z)\Pro_Z(z). 
\label{eq:ETbound}
\end{equation}


\nocite{*}
\bibliography{wileyNJD-AMA}%

\end{document}